# Guided flux motion: models and experiment

V.V. Guryev[1], S.V. Shavkin, V.S. Kruglov

*NRC Kurchatov institute, Kurchatov sq.1, 123182 Moscow, Russia*

The article provides a brief overview of existing models describing guided flux motion. The first model was proposed by Niessen et al. This model works in the single-vortex approximation and provides qualitative explanation of the effect, but in most cases it gives an overestimated guiding angle. The stochastic model explains the experimentally observed effect of decreasing the guiding angle, the so-called slipping effect, by the influence of thermal fluctuations. However, the performed estimates show that thermal fluctuations should not be so significant. An alternative guiding model is the anisotropic pinning model. This model works in the critical state approximation, and the slipping effect is the result of the combined action of anisotropic pinning and vortex interaction. The predictions of all three models are verified by the example of a superconducting Nb-Ti tape containing about 6 vol. % α-Ti, which acts as strong pinning centers. Cutting samples at different angles to rolling allows one to control the direction of the driving force. It was found that when the standard criterion of the electric field is reached, the guiding angle varies significantly at different places along the sample, which is explained by the plastic mode of the vortex matter motion. In this case, the anisotropic pinning model semi-quantitatively predicts the direction angle averaged over the sample length. With an increase in the driving force, the vortex matter motion passes into an elastic mode, and the guiding angle becomes uniform along the sample.

## I. INTRODUCTION

In light of the recent progress in quantum technology [1, 2] a field of science often called fluxonics [3, 4] acquires special relevance [5 – 8]. This name was given by analogy with electronics, since its basic element is a quantum of magnetic flux – fluxon. One of the key tools of the fluxonics is the guided flux motion (guiding) [4], a detailed description of which is given below. Generally speaking, the guiding effect is an essential attribute of the electrodynamics of any type II superconductor, including practical superconductors for high current applications. The guiding effect is determined by the quantum nature of superconductivity. This effect manifests itself as a non-coincidence of the electric field and current density directions and, so, contributes to the tensor character of the material equation of the superconductor [9], and consequently it should be taken into account in superconducting device designing.

Let us consider the case of a plate superconductor with a transport current density $\vec{j}$ in a perpendicular magnetic field $\vec{H}$. The magnetic field is assumed strong enough to satisfy the condition for the magnetic induction $\vec{B} \approx \mu_0 \vec{H}$. The current acts on the vortex matter with the driving force:

$$\vec{F} = [\vec{j} \times \vec{B}] \qquad (1)$$

This force leads to the movement of vortices with an average velocity $\vec{v}$, and, as a result, the emergence of an electric voltage $\vec{E}$ (Fig. 1a):

$$\vec{E} = [\vec{B} \times \vec{v}] \qquad (2)$$

Structural inhomogeneities impede the vortices motion. This is described by a pinning force opposite to the force (1). In the general case, when the defects are not globular and uniformly distributed, the pinning is anisotropic. Moreover, the anisotropy reveals itself not only in relation to the magnetic field direction $\vec{B}$, but also in relation to the driving force (1), which for a given direction of the magnetic field $\vec{B}$ is determined by the current direction $\vec{j}$. For example, pinning on a system of correlated planar grain boundaries is anisotropic (in both senses). In this case, the anisotropy with respect to force (1) is determined by the fact that the pinning force when vortices move along grain boundaries is much less than when they move across these boundaries. As a result, if the direction $\vec{F}$ is intermediate between these extreme cases (not along grain boundaries neither across them), then the average velocity of vortices $\vec{v}$ is not co-directed with the driving force (1): $\vec{v} \nparallel \vec{F}$. This situation is illustrated in Figure 1b. The electric field is still determined by (2), which leads to non-collinearity of the electric field and current density: $\vec{E} \nparallel \vec{j}$. This is how the component of the electric field transverse to the transport current $E_\perp$ appears. We define the guiding angle as:

$$\beta = \mathrm{atan}\frac{E_\perp}{E_\parallel} \qquad (3)$$

where $E_\perp$ and $E_\parallel$ - transverse and longitudinal to the transport current components of the electric field respectively.

---

[1] Corresponding author GuryevVV@mail.ru

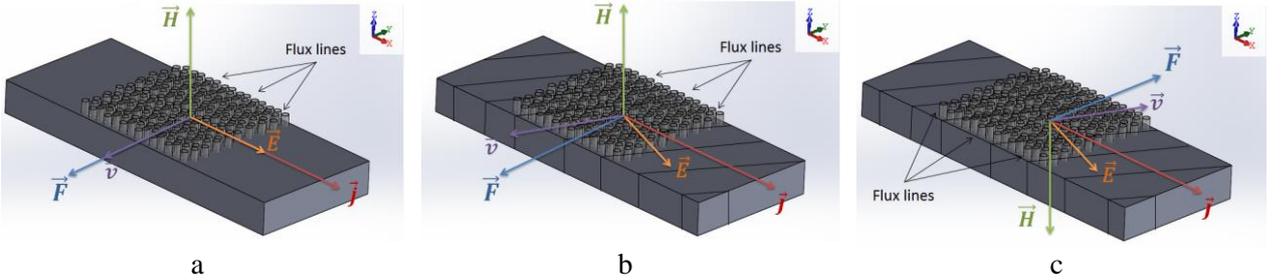

Fig. 1. Schematic representation of the vortices motion under the action of the driving force (1) (the magnetic field is applied normal to the tape): a) simple case when $\vec{F} \parallel \vec{v}$ ($\vec{j} \parallel \vec{E}$); b) the case when $\vec{F} \nparallel \vec{v}$; c) The similar to *b* case, but with the inverted direction of the magnetic field.

When the direction of the magnetic field is reversed, the direction of the electric field does not change (see Fig. 1c). Thus, the transverse component of the electric field $E_\perp$, caused by the guiding is even with respect to the magnetic field inversion. This important distinction from the Hall effect (which is odd with respect to the magnetic field inversion) allowed Nissen et al. experimentally detect this effect on rolled Nb sheets, cold-rolled Pb–In alloys [10] and Nb–Ta alloys [11] in the 60s. Already from these first experiments it became clear that the cause of the guiding is anisotropic pinning, but in most cases the direction of the vortex matter motion differs significantly from the direction of the easiest motion (along the grain boundaries) caused by the finite probability of the vortex movement across the grain boundaries (the so-called slipping effect). In [11], a model was proposed that qualitatively explains the observed features, which is hereinafter referred to as a naive model.

Below we consider the existing models explaining the features of the guided vortices motion (Section II), present the experimental results of the guiding-effect study (Section III) and compare the experimental results with theoretical models (Section IV).

## II. THEORETICAL MODELS

### A. Naive model

This model works in the so-called single-vortex approximation, i.e. the vortex is in the anisotropic pinning potential, but does not interact with other vortices. It is assumed that the vortex is sufficiently rigid and reacts to external excitation as a whole; therefore, particle-like equations of motion is used to describe the vortex dynamics. Then the balance of forces of a moving vortex is given as follows:

$$\vec{f_l} + \vec{f_p} + \vec{f_g} = \eta \vec{v} \qquad (4)$$

where $\vec{f_l} = [\vec{j} \times \vec{\phi_0}]$ – force acting on the vortex from the side of the transport current $\vec{j}$; $\vec{\phi_0}$ - vector along the magnetic field with a value equal to the magnetic flux quantum; $\eta$ – coefficient of viscous friction due to energy dissipation caused by the generation of an electric field; $\vec{f_p}$ – isotropic component of the pinning force directed opposite to the vortex velocity $\vec{v}$; $\vec{f_g}$ – anisotropic component of the pinning force perpendicular to the easy motion direction.

Let the sample be cut so that the direction of the transport current makes an angle $\xi$ with the direction of vortex easy motion (Fig. 2). For low enough current densities, the driving force $\vec{f_l}$ is small, and the perpendicular to $\vec{f_g}$ component $f_l \sin \xi$ can be fully balanced by $\vec{f_p}$. In this case, the vortex remains pinned. The beginning of the vortex motion is possible in the implementation of one of two scenarios.

1) If the force $\vec{f_g}$ is relatively large, then when $f_l \sin \xi$ exceeds the maximum value of $f_p$ the vortex begins to move in the direction of easy motion (Fig. 2.a). Then, with a further increase in $f_l$, when the component $f_l \cos \xi$ exceeds the maximum value of $f_g$, the vortex acquires a component of velocity perpendicular to the easy motion direction (slipping effect).

2) If the force $\vec{f_g}$ is relatively weak, then the vortex begins to move in the direction of the sum of forces $\vec{f_l} + \vec{f_g}$ (Fig. 2b). Accordingly, a further increase in the current will lead to a decrease in the guiding effect, since $\vec{f_g}$ is unchanged and $\vec{f_l}$ increases linearly with increasing current.

This model predicts that the choice between the two described scenarios is determined by the magnitude of the ratio of the maximum values $f_p/f_g$ compared to $\tan \xi$. At $f_p/f_g < \tan \xi$ the first scenario is realized, otherwise the second. The guiding angle $\beta$, at $\xi \neq 0°, 90°$ is defined as [11]:

$$\begin{cases} \beta = 90° - \xi, & \text{at complete guiding} \\ \beta = \operatorname{atan}[\frac{j}{j_s} \cos \xi \sin \xi - \frac{1}{\tan \xi}], & \text{at slipping effect} \end{cases} \quad (5)$$

where $j_s$ – current density at which the slipping effect begins.

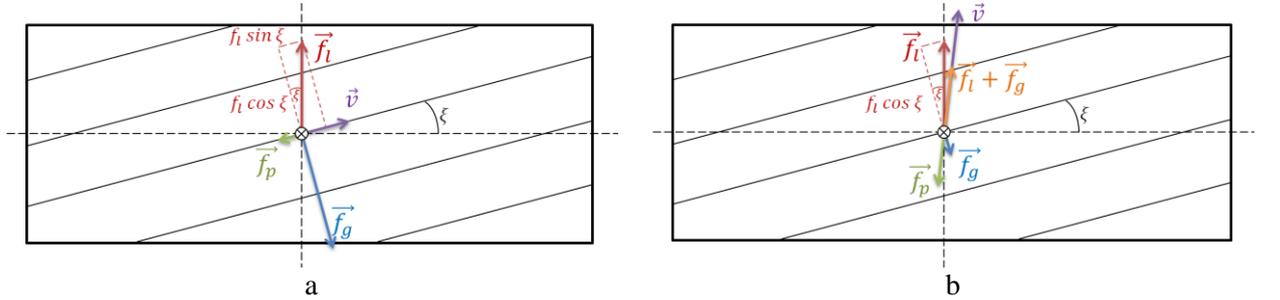

Fig. 2 Force diagrams in the model by Niessen et. al.: a) The case $f_p/f_g < \tan \xi$. The vortex begins to move along the direction of easy motion, b) The case $f_p/f_g > \tan \xi$. Immediately at the beginning of movement, the vortex has a velocity component perpendicular to the easy motion direction.

In the experiments of Niessen et al., the components $E_\parallel$ and $E_\perp$ of the electric field were measured depending on the value of the current, that is, the so-called two-dimensional current-voltage characteristics (2D-CVC). This made it possible to calculate the angle $\beta$ according to (3), which is equal to zero in the absence of the guiding effect. In order to be sure that the entire vortex matter is involved in motion, only parts of the CVCs which exceed the modern criteria for determining the critical current (~ 1 µV / cm) by 3-4 orders of magnitude in voltage were used. It was experimentally found that when the magnetic field $H$ changes in magnitude, the angle $\beta$ changes significantly. However, this phenomenon has not received a quantitative explanation within the framework of the naive model. In addition, the experiments have shown that slipping is a much more common phenomenon than the model predicts: the complete guiding, i.e. movement only along the easy motion direction is quite rare.

For a better fit with experimental data, Mawatari proposed to include thermal fluctuations into consideration through the use of the Fokker-Planck mathematical apparatus [12, 13]. This model was later developed in the works by Shklovskij et al. [14 – 18] and is called the stochastic model.

### B. Stochastic model

Here we will look at the general outline of the model. This model works in the same approximations as the naive model. The balance of forces of a moving vortex is written as:

$$\vec{f_l} + \vec{f_p} + \vec{f_g} + \vec{f_{th}} = \eta \vec{v} \tag{6}$$

in contrast to the naive model (4) $\vec{f_{th}}$ is added here - the force determined by thermal fluctuations. It is assumed that $\vec{f_{th}}$ is a stochastic force, i.e. random, with properties defined on average:

$$<f_{th,i}> = 0 \tag{7a}$$
$$<f_{th,n}(t)f_{th,m}(t')> = q\delta_{nm}\delta(t-t') \tag{7b}$$

where $q$ – thermal noise magnitude, $<...>$ - statistical averaging, $i, n, m = x, y$ – force components $\vec{f_{th}}$, $t$ – time, $\delta(...)$ – delta function, $\delta_{nm}$ - Kronecker symbol (1 if $n = m$, 0 if $n \neq m$).

Condition (7a) means that, on average, the force of thermal fluctuations is zero, while condition (7b) means that there are no correlations between different components of the force and the magnitudes of the forces at different moments of time, i.e. fluctuations are white noise. The quantity $q$ is assumed to be defined by temperature $T$.

Further, taking into account the stochastic nature of the force $\vec{f_{th}}$, one can proceed to a probabilistic description of the vortex dynamics within the framework of the Fokker - Planck equation:

$$\eta \vec{S} = (\vec{f_l} + \vec{f_p} + \vec{f_g})P - k_B T \nabla P \tag{8}$$

where $k_B$ – Boltzmann constant. Equation (8) operates with such quantities as $P(\vec{r}, t)$ – the probability density to find the vortex at time $t$ at point $\vec{r} = (x, y)$ and $\vec{S}(\vec{r}, t) \equiv P(\vec{r}, t)\vec{v}(\vec{r}, t)$ – vortex probability flux density. These quantities are related as:

$$\frac{\partial P}{\partial t} = -\nabla \vec{S} \tag{9}$$

The average vortex velocity $<\vec{v}>$, which, in accordance with (2) determines the electric field, can be found by:

$$<\vec{v}> = \frac{\iint \vec{S} d^2\vec{r}}{\iint P d^2\vec{r}} \tag{10}$$

The last term in Eq. (8) has a clear physical meaning: the temperature levels out the gradient of vortex concentration, and this leveling is the more intense, the higher the temperature. The real influence of this effect can be estimated by comparing the thermal energy $k_B T$ with the energy gain from the vortex pinning effect:

$$u_p \approx r_p f_p \tag{11}$$

here $r_p$ - the typical vortex displacement at which the pinning center ceases to act on the vortex.

In most cases, for sub-nitrogen temperatures $\frac{k_B T}{u_p} \sim 10^{-2} \cdots 10^{-6}$. This casts doubt on the adequacy of the temperature explanation for the slipping effect. In the work [16] Shklovskij et al. note that, apparently, the stochastic model is not applicable for low-temperature superconductors, leaving, however, a certain narrow region $(T_c - T) \ll T_c$ in which this approach may still be legitimate. However, in this region, the electrodynamics is determined mostly by inhomogeneity [19], up to the absence of vortices in this region as such [20 – 24]. A detailed review of the objections to the existing paradigm about the decisive influence of thermal fluctuations in the electrodynamics of practical superconductors can be found in the appendix to the work [25].

If one rejects the hypothesis of thermal fluctuations, the question remains: what makes the vortices move perpendicular to the easy motion? It is noteworthy that both the naive model and the stochastic model use the single-vortex approximation. On the other hand, the strong dependence of the guiding angle on the magnetic field, which was discovered already in the first works, indicates the key role of vortex interaction. Some attempts have been made to take this interaction into account within the framework of the dynamics of a single vortex by adding into (4) or (6) the force $\vec{f}_{vv}^i$ acting on the i-th vortex from the rest (see for example [26, 27]). However, perhaps due to too many independent fitting parameters, it was not possible to obtain any qualitatively new results in comparison with the already available ones. Even the problem of calculating the bulk pinning force from known pins concentrations and individual pinning forces is a difficult, still unsolved in the general case problem, known as the summation problem [28]. The variety of dynamic modes in the presence of correlated pinning centers [29] makes this approach even more laborious.

At about the same time as the work of Mawatari [12], an alternative approach to the electrodynamics of superconductors was proposed, taking into account the vortex interaction, but avoiding the summation problem [9]. Below we describe the general outline of this model. For a detailed acquaintance with this model, one can refer to the original work [9] or to the review [30].

### C. Anisotropic Pinning Model

The anisotropic pinning model works in the critical state approximation. An ensemble of a sufficiently large number of vortices is considered as a certain whole located in a cooperative potential well caused by the entire ensemble of pinning centers at once. In the absence of the transport current, this ensemble of vortices occupies the most advantageous place at the bottom of the potential well. The minimum of the specific energy at rest does not imply that each of the vortices captured by individual pinning centers is at the bottom of its individual potential well. Under the action of a transport current, an ensemble of vortices is jostled along the slope of the well. If the Lorentz force turns out to be greater than the maximum slope angle of the well, the vortices start to move. This leads to the appearance of an electric field and energy dissipation. Let's define the maximum pinning force opposing (1) as:

$$\vec{F}_p = -\max\left(\frac{\partial U}{\partial \vec{l}}\right) = -\vec{e}_l \frac{U_0(\vec{B})}{L_0(\vec{J}, B)} \tag{12}$$

where $U$ – the depth of the cooperative potential well, $\vec{e}_l$ – is the unit vector in the direction of the driving force (1), $L_0(\vec{J}, B) = \frac{U_0(\vec{B})}{|\max\left(\frac{\partial U}{\partial \vec{l}}\right)|}$ is the effective size of the cooperative potential well, $U_0(\vec{B})$ — the effective depth of the cooperative potential well.

An essential assumption of the model is that the effective depth of the cooperative potential well $U_0(\vec{B})$, is assumed to depend only on magnetic induction $\vec{B}$ (and does not depend on the current density $\vec{J}$ and, therefore, the driving force (1)), and the size of the cooperative potential well $L_0(\vec{J}, B)$ assumed to depend on the absolute value of the magnetic induction (density of the vortices) and the current density $\vec{J}$ (therefore, it depends on the driving force (1)) but does not depend on the direction of magnetic induction.

According to this model all pinning features in a particular material are determined by the specific angular dependences of the depth $U_0$ and the size $L_0$. In the case of Nb-Ti tape, the extreme values of $U_0$ and $L_0$ are achieved with orientations of magnetic induction and driving force along the main orthogonal directions in the material: normal direction (ND), rolling direction (RD) and direction perpendicular to rolling in the plane of the tape (PD). For other directions, $U_0$ and $L_0$ smoothly vary from minimum to maximum. Thus, the depth $U_0$ and the width $L_0$ are described by ellipsoids [9]:

$$\left(\frac{\cos\alpha''}{U^{RD}}\right)^2 + \left(\frac{\cos\beta''}{U^{PD}}\right)^2 + \left(\frac{\cos\gamma''}{U^{ND}}\right)^2 = \frac{1}{U_0^2} \tag{13a}$$

$$\left(\frac{\cos\alpha'}{L^{RD}}\right)^2 + \left(\frac{\cos\beta'}{L^{PD}}\right)^2 + \left(\frac{\cos\gamma'}{L^{ND}}\right)^2 = \frac{1}{L_0^2} \tag{13b}$$

where $\cos\alpha''$, $\cos\beta''$, $\cos\gamma''$ – are the direction cosines of induction $\vec{B}$; $\cos\alpha'$, $\cos\beta'$, $\cos\gamma'$ are the direction cosines of the driving force vector; $U^{RD}$, $U^{PD}$, $U^{ND}$, $L^{RD}$, $L^{PD}$, $L^{ND}$ depth and width of the cooperative potential well in the main orthogonal directions.

Let the external magnetic field be normal to the superconductor plane so $\vec{B} = (0,0,B)$ (Fig. 1). Then, in accordance with (13a) $U_0 = U^{ND}$. This direction of induction corresponds to the two-dimensional section of the ellipsoid $L_0$, which is determined by all possible directions of the force $\vec{F}$. As can be seen from (1) with a fixed direction of $\vec{B}$, the direction of the force $\vec{F}$ is determined by the current density direction. Thus, we have obtained the conditions for the angles of the direction cosines $\gamma' = 90°$ and $\beta' = 90° - \alpha'$, so from (13b) we get:

$$L_0(\alpha') = \frac{L^{RD}L^{PD}}{\sqrt{(L^{RD}\sin\alpha')^2 + (L^{PD}\cos\alpha')^2}} \quad (14)$$

Then, in accordance with (12):

$$(F_p(\alpha'))^2 = \left(\frac{U^{ND}}{L^{RD}}\right)^2 \cos^2\alpha' + \left(\frac{U^{ND}}{L^{PD}}\right)^2 \sin^2\alpha' \quad (15)$$

Figure 3 shows the angular dependence of the pinning force (15) and gives an explanation of the guiding effect. Let the direction of the transport current in the material be such that the driving force (1) is directed along the segment **ac** and makes an angle $\xi$ with the PD-direction. In the case of anisotropic pinning the values of the pinning force are different for different directions; there are directions for which the magnitude of the pinning force is so small that the projection of the force (1) on this direction reaches a critical value earlier (at lower currents) than in the direction of the force (1). This corresponds to the achievement of the value **ab** in the **ac** direction (Fig. 3): the projection **ad** reaches the critical pinning value, while the critical value has not yet been reached in the **ac** direction (**ab** < **ac**). As a result, the fluxoids flow not along the action of driving force (1), but along the critical projection of this force, which ultimately leads to the observed deviation of the electric field vector from the direction of the current by an angle $\beta$. In fig. 3, the dashed line shows the calculated locus of points corresponding to the appearance of the first such critical projection in the model of a double-ellipsoidal potential well. This curve is an ellipse with principal radii $\frac{U^{ND}}{L^{RD}}$ and $\frac{U^{ND}}{L^{PD}}$ [9].

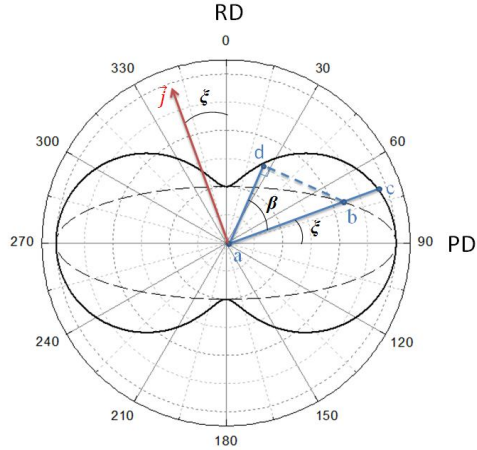

Fig. 3 Diagram of the guiding effect origin in the framework of anisotropic pinning model. The solid line corresponds to anisotropic pinning (14) described by two ellipsoids (13). The dotted ellipse corresponds to the appearance of the first critical projection when the driving force (1) increases from zero.

Taking into account that $\alpha' = 90^0 - \xi$, we get:

$$F_p(\xi) = F_p^{PD}\sqrt{\cos^2\xi + \gamma^2 \sin^2\xi} \quad (16)$$

where $\gamma = \frac{L^{RD}}{L^{PD}} = \frac{F_p^{PD}}{F_p^{RD}}$ – anisotropy factor determining the guiding angle $\beta$:

$$\beta = \text{atan}\left[\frac{\tan\xi}{\gamma^2}\right] - \xi \quad (17)$$

Thus, in this model, the effect of the guided flux flow is the result of the combined action of anisotropic pinning and inter-vortex interaction. The latter is taken into account since the factor $\gamma$ is allowed to vary depending on the magnitude of the magnetic field. Note that the model does not require an assumption about significant thermal fluctuations influence. The low-field guiding angle predicted by this model was recently confirmed on a cold-rolled Nb-Ti tape using the magneto-optical studies [31].

A similar geometric approach for interpreting the guided flux motion was also proposed in later works by Brandat and Mikitik [32, 33], however, the first assumption that the pinning force is determined not by two ellipsoids (13), but by one, did not allow the authors to achieve a plausible guiding angle. For better agreement with experiment, the authors then modernized the model to describe the angular dependence of the pinning force by a formula similar to (14) [33], however, they did not explain the origin of such a form. In the anisotropic pinning model, the fundamental need for two ellipsoids is due to the fact that, as noted in the introduction, the anisotropy of the pinning force is of two types: i) with respect to the direction of the magnetic field at a fixed direction of driving force (1), ii) with respect to the driving force (1) at a fixed direction of the magnetic field.

## 3 EXPERIMENTS
### A. Samples and experimental techniques

The samples were made of a wide (80 mm) Nb-Ti tape with a thickness of 10 microns. When choosing an Nb-Ti alloy as an object of research, we were guided mainly by two reasons. First, this is the most common practical superconductors with the well-developed production technology which allows one to guarantee uniformity of the properties and reproducibility of the results obtained on different samples. Secondly, the upper critical field of ~ 12 T (at 4.2 K) available for standard research equipment allows one to sweep the vortex system from the single vortex limit at low magnetic fields to the strongly interacting vortex limit at high magnetic field.

Compared with previously studied samples [9, 31], here we present the results for the Nb-Ti tape additionally heat-treated at 385 $^0$C for 25 hours. Previously, we studied in detail the microstructure of the tape, both by transmission electron microscopy [31] and by diffraction methods using laboratory facilities and synchrotron radiation [31, 34]. Relevant results about the structure of the tape are given below in this section.

In the cold rolled tape, the grain boundaries are the main pinning centers. The characteristic sizes of Nb-Ti grains are ~ 40 nm in the normal direction, 0.2 μm in the direction perpendicular to rolling, and more than 1 μm in the rolling direction. As a result of heat treatment, a significant change in the grain size was not observed, but ~ 6 vol.% of non-superconducting α-Ti particles, which are effective pinning centers, were precipitated. The size distribution of α-Ti particles is well approximated by a log-normal distribution with the most probable sizes of 16 nm in the normal direction (ND), 40 nm in the PD direction and 75 nm in the rolling direction. Such structural changes did not lead to a significant change in the values of the thermodynamic critical parameters ($T_c$ and $B_{c2}$) and their spread over the tape width (<1%) as compared to the cold-rolled tape. However, the width of the temperature transition increased by 1.5 times, which indicates an increase in micro inhomogeneity with unchanged macro inhomogeneity.

For presented work, samples were cut out with angles $\xi=0^0$, $30^0$, $60^0$ and $90^0$ to the rolling direction. To study the guided flux motion, a 6-contact transport method was used: 2 current contacts and 4 potential contacts soldered in two pairs. The contacts in each pair are located on opposite edges of the sample. This geometry makes it possible to control the lateral Hall-like electric field component at two different locations in the sample. The distance between the contacts in a pair is 3.5 mm, the distance between pairs along the sample is 8 mm. The error in the accuracy of the contacts arrangement in a pair was controlled by measurements in the normal state in the magnetic field higher then $H_{c2}$. The value of the mismatch of the transverse contacts obtained in this way was taken into account in further measurements. The measurement temperature was 4.2 K. The critical current $I_c$ was determined at 1 μV / cm.

The enhanced current-carrying capacity of the heat-treated samples as compared to the original cold-rolled ones led to the fact that samples were extremely unstable. For present study, the samples had a copper shunt layer. This, however, means that when a finite voltage appears across the superconductor, part of the current will be redistributed into the stabilizing layer. As a cut-off, above which this effect of current redistribution introduces an unacceptably large distortion, the value of reaching 10% of the resistance in the normal state, measured at a field above the upper critical, was taken.

### B. Results

Figure 4 shows the field dependences of the volume pinning force, defined as $F_p = j_c \mu_0 H$ for samples cut at different angles to rolling. For comparison, the same figure shows the field dependences of the pinning force for the cold-rolled tape in two cases: $\xi = 0^0$ and $90^0$. All $F_p(H)$ dependences strongly differ from the dome-shaped one, which is typical for strong pinning. Compared to cold-rolled one the pinning in the heat-treated tape in the middle range of fields ($0.5\,T < \mu_0 H < 8.5\,T$) increased for all samples. At that the double-humped shape of the $F_p(H)$ dependence characteristic for the cold-rolled tape, although preserved, became less pronounced in case of heat-treated tape. This behavior is consistent with the known data on the enhancement of pinning during heat treatment [35, 36]. The appearance of two peaks is often interpreted as a field-induced crossover from the strong to the weak regimes of pinning. The theory of such crossovers was developed in the classical work by Larkin and Ovchinnikov [37]. At low ($\mu_0 H < 1\,T$) and high ($\mu_0 H > 9\,T$) fields, the vortex lattice is deformed plastically, which leads to a strong pinning (plastic static state), while in an intermediate field the elasticity of the vortex lattice dominates over the pinning force (elastic static state), resulting in the weak collective pinning.

After heat treatment, not only the increase in the pinning force is observed, but also a significant decrease in the anisotropy with respect to the driving forces. Figure 5 shows the field dependence of the anisotropy factor which is defined as a ratio for cases when the sample is cut across ($\xi = 90^0$) and along ($\xi = 0^0$) to the rolling direction $\gamma(H) = \dfrac{F_p^{\xi=90}}{F_p^{\xi=0}}$.

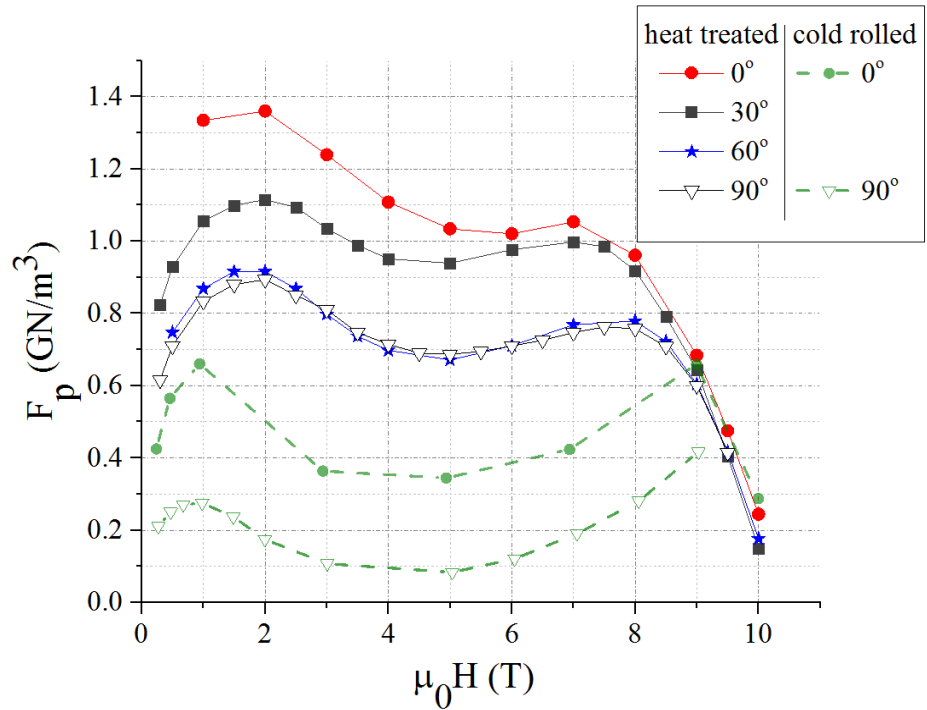

Fig. 4 Field dependence of the pinning force of the heat-treated Nb-Ti tape for samples cut at different angles to the rolling direction. The field is directed normal to the tape. For comparison, the field dependences of a cold-rolled tape are also shown.

Starting from the critical current value $j_c$, a detectable experimentally electric field appears. With a further increase in the current from the experimental 2D-CVC, it is possible to extract the guiding angle according to (3). Figures 6 and 7 show the guiding angle dependencies on the transport current for samples cut at angles 30° and 60° respectively. The two solid lines correspond to the two pairs of transverse voltage potential contacts as described in Section 3.A. Registration of the $\beta(I)$ dependence ended either with a violation of the *I*-V characteristic due to the magnetic flux jumps, or reaching 10% of the normal resistance. At low currents, a high level of noise is observed, since in this region the voltage differs only slightly from zero, which creates difficulties in experimental detection.

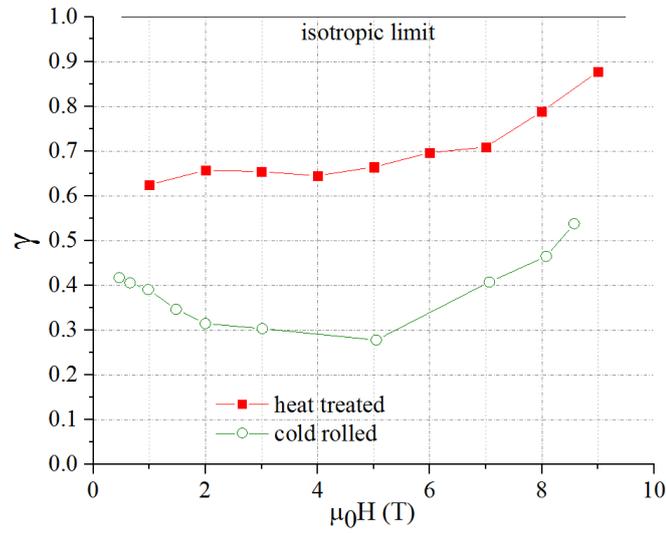

Fig. 5 Pinning anisotropy with respect to the direction of the driving force defined as $\gamma = \dfrac{F_p^{\xi=90}}{F_p^{\xi=0}}$

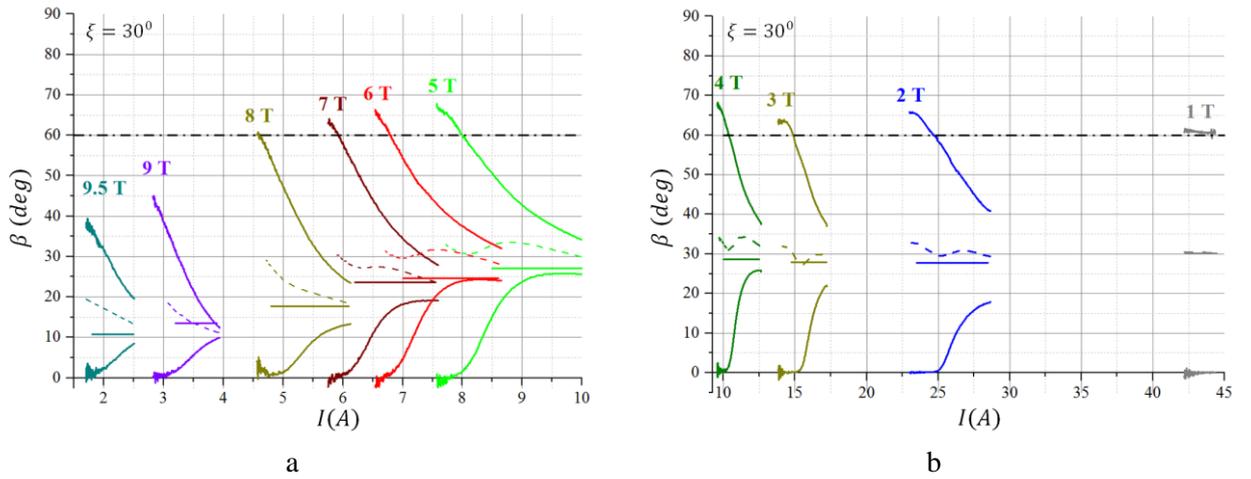

Fig. 6 Guiding angle $\beta(I) = \operatorname{atan}\dfrac{E_\perp(I)}{E_\parallel(I)}$ for a sample cut at an angle $\xi = 30°$ t different magnetic fields: a) 9.5 T – 5 T; b) 4 T – 1 T. he solid lines show the experimental data for two transverse pairs of contacts, the dashed line is the averaging of these two pairs. Horizontal lines are model predictions (17), horizontal dash-and-dot lines correspond to full guiding.

For the convenience, Figures 6 and 7 also show the angles corresponding to the full guiding, i.e. no slipping effect (black dash-dot), and guiding angles expected in the AMP model (solid horizontal lines). Note that in all experiments the curves in Figures 6 and 7 are fully reproduced with an increase and decrease in the transport current, which indicates that the magnetic flux in both cases moves along the same trajectory.

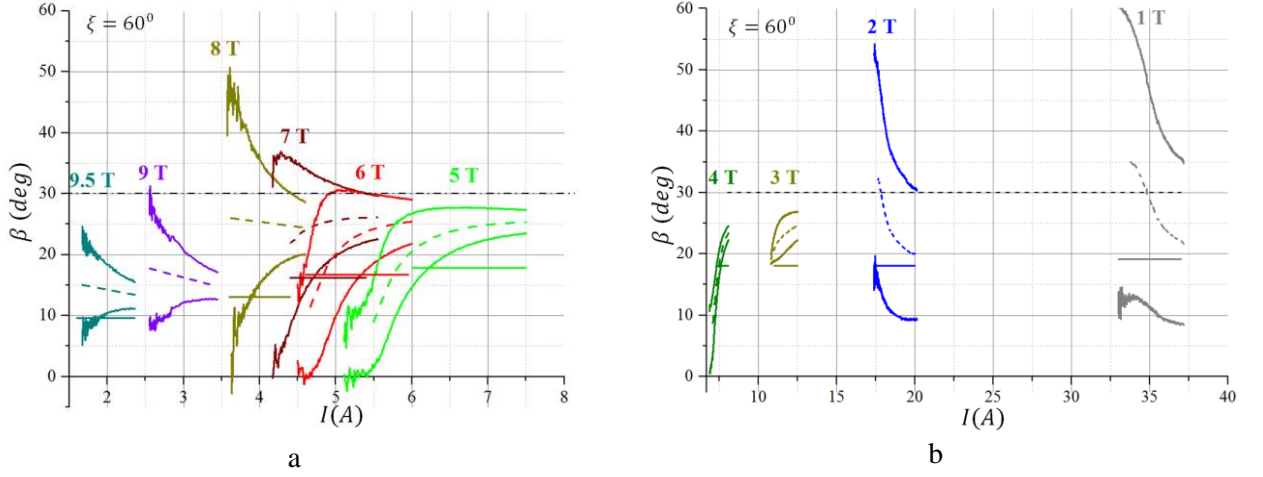

Fig. 7 Guiding angle $\beta(I) = \operatorname{atan}\frac{E_\perp(I)}{E_\parallel(I)}$ for a sample cut at an angle $\xi = 60°$ t different magnetic fields: a) 9.5 T – 5 T; b) 4 T – 1 T. he solid lines show the experimental data for two transverse pairs of contacts, the dashed line is the averaging of these two pairs. Horizontal lines are model predictions (17), horizontal dash-and-dot lines correspond to full guiding..

## 4 DISCUSSIONS

In low fields, when the vortex interaction is small, it can be assumed that the anisotropy factor $\gamma \approx \frac{f_p}{f_p+f_g}$, where $f_p$ and $f_g$ – isotropic and anisotropic components of pinning in accordance with the equation (4) of the naive guiding model. Thus, the value of the ratio $f_p/f_g \approx 1.5$. This ratio is greater than $\tan(30°) = 0.577$, but less than $\tan(60°) = 1.732$, and therefore in accordance with the predictions of this model for the angle $\xi = 30°$ the slipping effect is realized, while at the angle $\xi = 60°$ the compete guiding should be observed. However, it is clearly seen from Figures 6 and 7 that complete guiding is not realized either for the angle $\xi = 30°$ nor for $\xi = 60°$.

The estimates made within the framework of different models for the elementary pinning energy of the cold-rolled tape agree well with each other and give the value $u_p \approx 2 \cdot 10^{-21} J$ [38]. Since the characteristic grain sizes did not change after heat treatment, and the bulk pinning increased, as can be seen from Figure 4, the value of elementary pinning energy is apparently even higher for the heat-treated tape. Hence the value $u_p \approx 2 \cdot 10^{-21} J$ can be used as a lower estimate for individual pinning in the heat-treated tape. Thus the ratio $\frac{k_B T}{u_p}$ is not more than 3% and it is not possible to explain the pronounced slipping effect by thermal fluctuations.

The anisotropic pinning model gives the most accurate predictions for the guiding angle and its dependence on the magnetic field, which can be considered as semi-quantitative (Fig. 6 and Fig. 7). The critical state approximation is justified for the cold rolled tape with weak pinning, since it did not show a significant dependence of the guiding angle on the current value [9]. On the contrary, for the heat-treated tape at low values of the electric field (near the criterion $E_c = 10^{-4} V/m$), the guiding angle is not the same in different places of the tape, which indicates the plastic nature of the vortex matter motion, i.e. vortices can exchange neighbors while moving. In other words, due to the inhomogeneity of the pinning landscape, percolation paths appear in the vortex matter, so some chains of vortices are moving while the rest of the vortex matter is motionless [39, 40]. Thus, the heat treatment leads to an increase in pinning strength and a crossover from elastic to plastic flow, which is also consistent with well-known trends [41].

At higher drives, the effective pinning forces experienced by the moving vortex matter become weak enough that the vortex matter can dynamically form a moving elastic solid [42]. This corresponds to the tendency for the two branches to meet in Figures 6 and 7 at high currents.

The above reasoning explains why, in the plastic flow mode, the prediction of the guiding angle in the APM model can be correct, only on average, while the elastic flow regime is much better in agreement with the experiment.

To conclude this section, we would like to note that several simulations have been performed to shed light on the influence of the plastic-elastic flow crossover on the guiding effect [42, 43]. The obtained experimental data can serve as a guideline for further work on this kind.

## CONCLUSION

The models of guided flux motion and their limitations are considered. It is shown that in the case of low-temperature superconductors the anisotropic pinning model (APM) is the closest to the experimental observations. In the framework of the APM model, the cause of the slipping effect is the result of the combined action of the vortex interaction and the anisotropy of the pinning centers, while, a hypothesis about the significant influence of thermal fluctuations is not required. This model semi-quantitatively describes the field dependence of the guiding angle in Nb-Ti tape in case of weak collective pinning, as well as in the case of strong pinning and the elastic mode of the vortex matter motion, which is realized at high enough driving force. In the case of strong pinning and plastic flow mode, this model predicts the guiding angle averaged over the sample length.